# Generation of energetic electrons by surface waves in VHF CCPs

D. Eremin [1], E. Kemaneci [1], M. Matsukuma [2], I. Kaganovich[3], T. Mussenbrock [1], R.P. Brinkmann [1]

[1] *Ruhr University Bochum, Bochum, Germany*
[2] *Tokyo Electron Technology Solutions Limited, Nirasaki City, Japan*
[3] *Princeton Plasma Physics Lab, Princeton, New Jersey, USA*

Capacitively coupled plasmas (CCP) comprise one of the main tools in active use in the plasma processing industry. However, increasing the driving frequency and electrode size is limited by the emergence of plasma radial nonuniformity detrimental to applications. The nonuniformity is caused by interactions of surface waves natural to the plasma-filled reactor [1,2] with electrons of the plasma, leading to complex electron energization and ionization dynamics. To demonstrate this, we use the fully electromagnetic 2d3v particle-in-cell GPU-parallelized code ECCOPIC2M, described in detail in [3], to simulate the "Testbench B" experiment conducted previously [4].

The code utilizes the implicit charge- and energy-conserving algorithm initially suggested in [5] and adapted for bounded collisional plasmas in [6], then generalized further to 2d cylindrical geometry [7] and a fully electromagnetic model [3]. Such a numerical algorithm does not suffer from the usual limitations on the time step and the cell size (which have to be smaller than the light-wave transit time through a cell and the Debye length, respectively), plaguing the conventional explicit momentum-conserving PIC algorithm. This significantly reduces the net computation time, despite the increased algorithmic complexity.

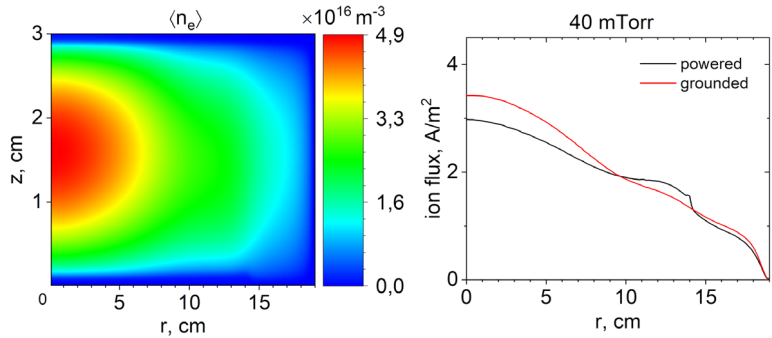

**Fig.1**: Time-averaged electron density and ion flux radial profiles at the powered and grounded electrodes obtained from a simulation of a CCP discharge in argon at 40 mTorr, driven at 106 MHz and 37 W.

The simulated geometry models an (r-z) cross-section of the cylindrical capacitively coupled reactor. Fig.1 shows the reactor chamber with plasma, which is bounded at the bottom by a powered electrode (r<14 cm) and by a dielectric spacer (14 cm < r < 19 cm), at the side by a grounded wall, and at the top by a grounded electrode. One can see that for a discharge driven at 106 MHz, as indicated in the caption to Fig.1, plasma density has indeed a strongly nonuniform radial profile. Ion flux at the electrodes, which is more relevant for the plasma processing technologies, also exhibits a radial nonuniformity, whereas the ion energy distribution is virtually radially uniform at the electrodes (not shown here; for details, see [3]), which is to be expected for conducting electrodes. As demonstrated in [3], this radial nonuniformity is caused by the radial nonuniformity of the ionization source.

Since the ionization is caused by energetic electrons with kinetic energy above the ionization threshold, the ionization rate profile is intimately related to the electron energization, i.e., the generation of such fast electrons. On the one hand, this can occur as a result of the collision-dominated Ohmic heating caused by the relatively small energy gains from the electric field on a distance approximately equal to the mean free path. Formation of the energetic electron tail results then from collisional diffusion in energy space. On the other hand, there is an essentially collision-less electron heating mechanism enabling electrons to gain large energy through interaction with the potential barrier of an expanding plasma sheath [8]. Such a mechanism is referred to as the "stochastic" or "pressure" heating and can be enhanced by excitation of the plasma series resonance (PSR) by the sheath motion [9] or by collisions reversing the electron motion, which cause additional collisions with the expanding sheath and lead to further increase in the kinetic energy [10]. At low pressures, the second mechanism can become comparable to or even dominate over the first one.

There are two important aspects related to electron heating, the first being the total energy absorbed by electrons and the second being the energy absorbed per electron. Fig.2, left, shows radial profiles of the total power density absorbed by electrons from the electric field in the bottom (dashed) and the top (dotted) half of the discharge, as well as their sum (solid line).

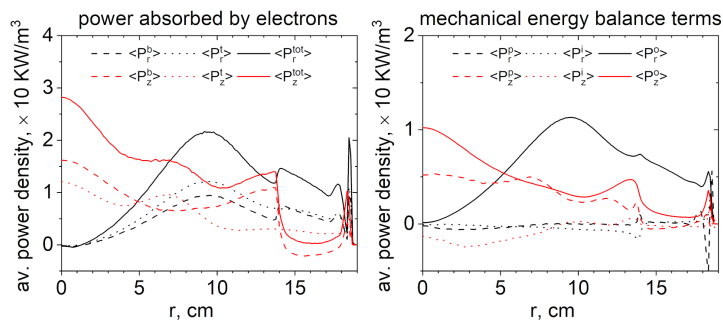

**Fig.2**: Time-averaged power density absorbed in the radial (black) and axial (red) directions in the top and bottom half of the discharge (left) and its "mechanical energy balance" decomposition [11] (right).

It can be observed that the biggest difference is demonstrated by the radial profiles of the total power density absorbed by electrons from the radial and the axial electric field. The former peaks close to r = 10 cm and is small close to the center, while the latter peaks at the center and falls down to the edge. Recalling the centrally peaked radial profiles of the electron density and the ion flux, one can conclude that it is the electron heating in the axial direction that is responsible for the observed radial nonuniformities. One can note that electrons are similarly heated in different discharge halves, except the region above the dielectric spacer and the difference between the power density absorbed in the axial direction close to the powered electrode's edge (9 cm < r < 14 cm).

Fig.2, right, plots different contributions in the "mechanical energy balance" analysis [11]: the pressure, the inertial, and the Ohmic terms. Evidently, electron heating in the radial direction is by far dominated by the Ohmic term, whereas in the axial direction, both pressure and Ohmic terms are significant. The difference can be attributed to the various heating mechanisms: whereas in the radial direction, it is mostly of the Ohmic nature and reflects the profile of the fundamental radial eigenfunction [1,2], in the axial direction, the heating occurs to a significant degree due to the interaction with the electric field of the expanding sheath, which accelerates electrons towards the bulk. As argued in [3], such an interaction energizing electrons often occurs predominantly at the fundamental harmonic of the driving frequency if the latter is not very high and the driving field amplitude is not very large so that no significant excitation of higher harmonics happens. Otherwise, higher harmonics can substantially contribute to the generation of energetic electrons.

As far as the electron energization is concerned, it is more important to know how much energy is absorbed per electron rather than the total energy absorbed by the whole electron population. Fig.3 shows that, despite comparable total absorbed power at the center and the edge, a lot more energetic electrons with energy above the ionization threshold are produced close to the center. Again, this can be attributed to the very high efficiency of the electron energization due to the axial sheath expansion, which can be enhanced due to the excitation of higher surface modes and due to the fact that a surface wave, which is excited at a periphery and propagates towards the center, grows in amplitude due to the energy conservation.

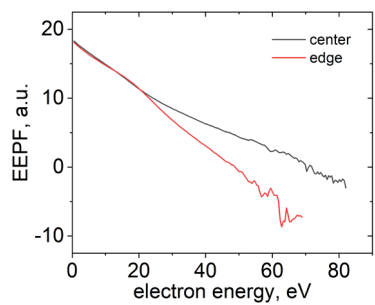

**Fig.3**: Electron energy probability function obtained from the described simulation, sampled at the discharge center (black) and the edge of the powered electrode (red).

An indirect indication of such a phenomenon is observed in Fig.4, where a population of energetic electrons is produced as a result of the surface wave excitation at r = 8 cm. It can be seen that the density of energetic electrons is higher at the center. One can also see how the electrons are successively produced as the corresponding surface wave propagates toward the center. The resulting beam of energetic electrons propagates axially towards the bulk, taking part in the ionization events and only weakly decreasing in intensity. The reason is that the electron mean free path in Ar at 40 mTorr and 10 eV is approximately 3 cm, which is equal to the electrode distance.

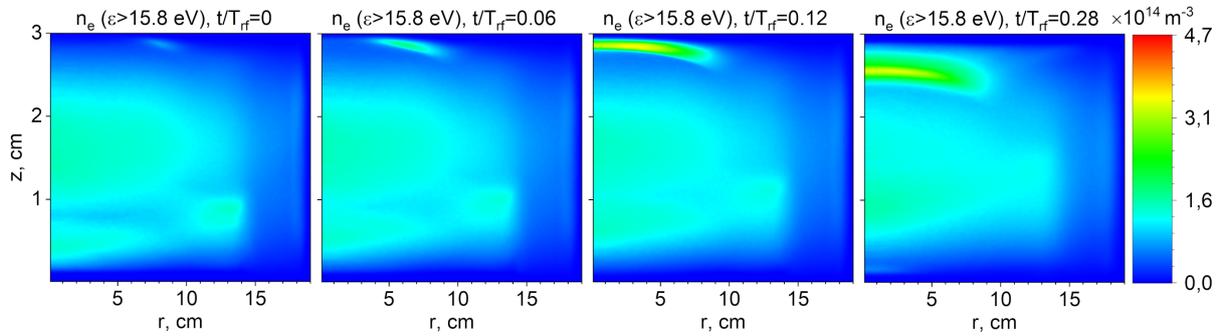

**Fig.4**: Evolution of an energetic electron population at the grounded electrode as a result of the excitation of surface wave at r = 8 cm, shortly before t=0.

One can see that the underlying physics involves an interplay of different phenomena and requires a self-consistent, kinetic, nonlocal, and electromagnetic description, which takes into account the finite electron inertia effects. We believe that the implicit energy-conserving electromagnetic PIC/MCC code utilized in this study fully meets all these requirements.

**References:**

[1] M.A. Lieberman *et al* 2002 *Plasma Sources Sci. Technol.* **11** 283

[2] L. Sansonnens *et al* 2006 *Plasma Sources Sci. Technol.* **15** 302

[3] D. Eremin *et al*, *Modeling of very high frequency large-electrode capacitively coupled plasmas with a fully electromagnetic particle-in-cell code,* arXiv:2212.08836v2 [physics.plasm-ph]

[4] I. Sawada *et al* 2014 *Jpn. J. Appl. Phys.* **53** 03DB01

[5] G. Chen *et al* 2011 *J. Comput. Phys.* **230** 7018

[6] D. Eremin 2022 *J. Comput. Phys.* **452** 110934

[7] D. Eremin *et al*, *Electron dynamics in planar radio frequency magnetron plasmas: II. Heating and energization mechanisms studied via a 2d3v particle-in-cell/Monte Carlo code,* arXiv:2211.03459 [physics.plasm-ph]

[8] M.M. Turner 2009 *J. Phys. D: Appl. Phys.* **42** 194008

[9] T. Mussenbrock 2008 *Phys. Rev. Lett.* **101** 085004

[10] J. Schulze *et al* 2015 *Plasma Sources Sci. Technol.* **24** 015019

[11] M. Surendra *et al* 1993 *Phys. Rev. E.* **48** 3914